# Orbital, Superhump, and Superorbital Periods in the Cataclysmic Variables AQ Mensae and IM Eridani

E. Armstrong[1, 2], J. Patterson[2], E. Michelsen[1], J. Thorstensen[3], H. Uthas[2], T. Vanmunster[4], F.-J. Hambsch[5], G. Roberts[6], S. Dvorak[7]
[1]*Department of Physics, University of CA, San Diego, La Jolla, CA 92037, USA*
[2]*Department of Astronomy, Columbia University, 550 West 120th St, New York, NY 10027, USA*
[3]*Department of Physics and Astronomy, Dartmouth College, Hanover, NH 03755-3528*
[4]*CBA-Belgium, Walhostraat 1A, B-3401 Landen, Belgium*
[5]*CBA-Antwerp, Vereniging Voor Sterrenkunde, Oude Bleken 12, B-2400 Mol, Belgium*
[6]*CBA-Tennessee, 2007 Cedarmont Drive, Franklin, TN, 37067, USA*
[7]*CBA-Orlando, Rolling Hills Observatory, 1643 Nightfall Drive, Clermont, FL 34711, USA*

## ABSTRACT

We report photometric detections of orbital and superorbital signals, and negative orbital sidebands, in the light curves of the nova-like cataclysmic variables AQ Mensae and IM Eridani. The frequencies of the orbital, superorbital, and sideband signals are 7.0686 (3), 0.263 (3), and 7.332 (3) cycles per day (c $d^{-1}$) in AQ Mensae, and 6.870 (1), 0.354 (7), and 7.226 (1) c $d^{-1}$ in IM Eridani. We also find a spectroscopic orbital frequency in IM Eridani of 6.86649 (2) c $d^{-1}$. These observations can be reproduced by invoking an accretion disc that is tilted with respect to the orbital plane. This model works well for X-ray binaries, in which irradiation by a primary neutron star can account for the disc's tilt. A likely tilt mechanism has yet to be identified in CVs, yet the growing collection of observational evidence indicates that the phenomenon of tilt is indeed at work in this class of object. The results presented in this paper bring the number of CVs known to display signals associated with retrograde disc precession to twelve.

We also find AQ Men to be an eclipsing system. The eclipse depths are highly variable, which suggests that the eclipses are grazing. This finding raises the possibility of probing variations in disc tilt by studying systematic variations in the eclipse profile.

## 1. INTRODUCTION

**"The ringing became more distinct ... it continued and gained definitiveness ... It was a *low, dull ... sound – much such as a watch makes when enveloped in cotton*."**

Edgar Allan Poe; *The Tell-tale Heart* (1843)

Superorbital periods associated with accretion disc precession have been reported in well-studied low-mass X-ray binaries (LMXBs) since the first identification of this phenomenon in Hercules X-1 (Katz 1973; Roberts 1974; Petterson 1977). These periods are attributed to the motion of the accretion disc, in which different annuli coordinate their rotation rates to effect one of two types of precession: 1) prograde precession of the line of apsides of an eccentric disc, and 2) retrograde precession of the line of nodes of a disc that has tilted out of the orbital plane. It is the latter of these two types that is discussed in this paper.

Superorbital periods associated with retrograde disc precession are also observed in cataclysmic variables (CVs), although not as commonly: prior to the publication of this paper, only about ten CVs had been known to display them. This small sample size is due at least in part to the relatively few searches that have been conducted for these signals in CVs. Firm detections of such long-period signals require observations baselines of at least several weeks, for sufficient frequency resolution and for discrimination among the daily aliases that afflict ground-based photometry.

We report detections, in the nova-like CVs AQ Mensae (AQ Men) and IM Eridani (IM Eri), of orbital, superorbital, and negative orbital sideband periods (hereafter “negative superhump periods”) obtained via long-term time series photometry. We also report a spectroscopic orbital period for IM Eri. Given the presence of the two faster signals, it is likely that the superorbital signals are due to retrograde disc precession. We also establish AQ Men as an eclipsing CV, an identification that has been difficult to make, due to the high variability of the orbital waveform in the light curve. We shall argue that the variability is likely to arise from *grazing* eclipses of the disc (the disc being the dominant light source), which changes shape with respect to the observer as it precesses. AQ Men thus presents a potential laboratory for testing the model of tilted discs in CVs. That is: it may offer an opportunity to study the disc shape in detail, and as a function of the strengths and phases of other periodicities that are present in the light curve.

A short biography of each object appears in its respective section in this paper.

## 2. OBSERVATIONS and ANALYSIS

All data reported here were obtained by the globally distributed small telescopes of the Center for Backyard Astrophysics (see Skillman & Patterson (1993) for details of the CBA instrumentation and observing procedure). We obtained differential photometry of the CV with respect to a comparison star on the same field, and spliced overlapping data from different longitudes by adding small constants to establish a consistent instrumental scale. With an excellent span of longitudes, we essentially eliminated the possibility of daily aliasing of frequencies in the power spectra. In order to reach good signal-to-noise with good time resolution, we generally observe in unfiltered light. This practice, however, eliminates the possibility of transforming to a standard magnitude. The CCD detector has good response over 4000-9000 Ångstroms (Å), which yields an effective wavelength near 6000 Å for stars with colors typical of cataclysmic variables (B - V ~ 0). Thus the magnitudes correspond roughly to V light, although this cannot be made precise since the comparison star has a different effective wavelength (which varies with color and airmass).

Differential magnitudes were obtained by computing the light within circular apertures of ~ 5 – 10 arcsecond radius. Typical exposure times were about one minute, with five seconds of deadtime between images. After establishing a common magnitude scale, we inspected the spliced light curves for their general appearance. In this case, both stars' light curves looked typical of many nova-like variables. We searched the light curves for periodic signals by calculating Lomb-Scargle power spectra. Lomb-Scargle is similar to the discrete Fourier transform and works well for transforming data that are unevenly-spaced in time (Scargle 1982). Where warranted, we performed modeling and subtractions of periodic signals in the light curves via least-squares fitting of three-parameter waves (of frequency, amplitude, and phase). This method works well when the signal in question is approximately sinusoidal.

## 3. AQ MENSAE

AQ Mensae (AQ Men), previously EC0511-7955, was identified as a CV by Chen et al. (2001). Their spectroscopy showed broad double-peaked Balmer emission lines, with profile variations consistent with Doppler broadening in a disc. Radial velocities suggested an orbital period ($P_{orb}$) of 3.1 h (7.7 c $d^{-1}$). The photometry on 9 nights showed no outbursts and no orbital modulation, and mild variability with an average magnitude of B = 15.1, except for one night when they saw a 0.85-mag dip resembling an eclipse (although they did not suggest that eclipses were actually occurring). They tentatively classified the object as a dwarf nova, based on the large-amplitude flickering at low frequency and a spectrum that they described as one of fairly low excitation with features similar to those of known dwarf novae. But no outbursts or fadings were observed, and the spectrum - with both He II and the CIII/NIII blend in emission - is arguably more reminiscent of a nova-like variable. Thus those authors left open the question of proper classification. We favor the nova-like description because there is still no record of eruptions and the spectrum appears like that of a typical nova-like variable.

One other targeted observation has been made of AQ Men, obtained by the Far Ultraviolet Spectroscopic Explorer (Godon et al. 2009), the results of which were consistent with – but less constraining than – those of Chen et al. (2001). They also cited unpublished results of our photometry, which we now present properly in this paper.

### *3.1.1 Photometry and Analysis*

We observed AQ Men for 177 hours on 33 nights over 49 days in 2002; Table 1 contains the summary observing log. The top panel of Fig. 1 shows the full light curve; at bottom is a representative ~ 0.8-day window showing the variable modulation. The out-of-eclipse brightness remained relatively stable throughout observations, with a mean magnitude of V = 14.6 (2) mag, with a systematic (calibration) error of ~ 0.3 mag. The star showed rapid flickering and possibly one large dip near the end of the light curve in the lower frame.

The power spectrum of AQ Men over the 49 nights is shown in the top panel of Fig. 2, with features of interest noted. The strongest signal occurs at 0.263 (3) c $d^{-1}$ (inset at right). The other prominent features are five sharp peaks at frequencies that are exact multiples of 7.0686 (3) c $d^{-1}$, where the error was derived from the dispersion of the five frequency values. These features indicate a strong periodic signal at that fundamental frequency, but apparently with a very nonsinusoidal wave form (hence the many harmonics). When we performed a synchronous summation at 7.0686 c $d^{-1}$, we obtained the mean light curve shown in the lower frame. There is also a feature at 7.332 (3) c $d^{-1}$ (inset at left), which will be discussed in §3.1.2. Table 2 lists frequencies and amplitudes of the three main features of interest.

**Table 1**. Photometry

| *Object* | *Dates (UT)* | *Dates (JD: 2,450,000+)* | *Hours* |
|---|---|---|---|
| AQ Men | 2002 Feb 3 – Mar 24 | 2309-58 | 176.8 |
| IM Eri | 2002 Sep 20 – 1 | 2538-9 | 3.1 |
| | 2002 Nov 23 - 7 | 2602-6 | 38.4 |
| | 2002 Dec 14 - 22 | 2623-31 | 49.0 |
| | 2011 Dec 28 - 2012 Jan 29 | 5924-56 | 144.48 |

*Note: Observing windows averaged 3-5 hours per night.*

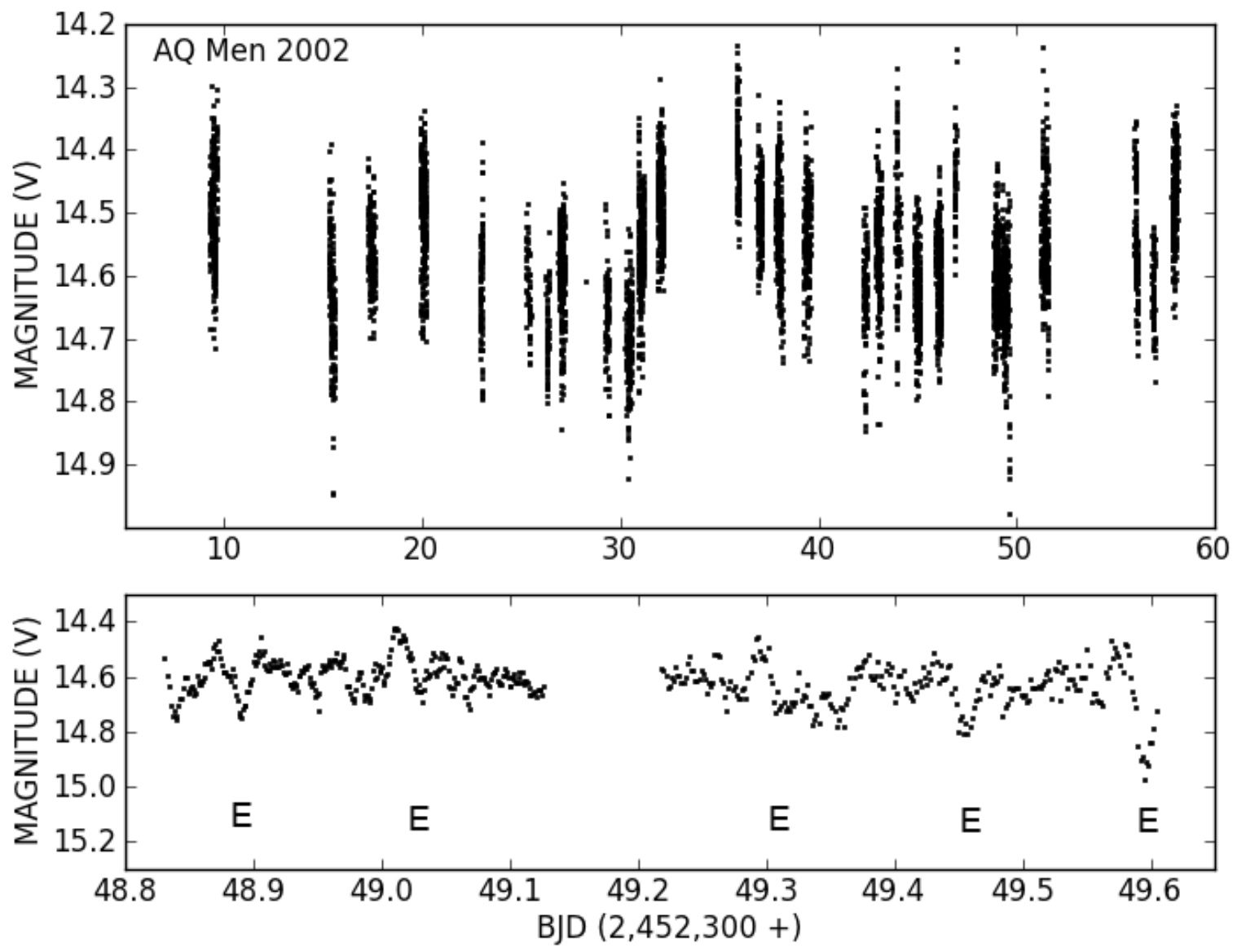

**Figure 1**. *Top panel:* Full light curve of AQ Men in 2002. *Bottom:* A 0.8-day subset of the light curve, illustrating the variable modulation. E's designate times of eclipses. The V magnitudes have a systematic calibration error of ~ 0.3 mag.

**Table 2**. Periodic Photometric Signals in AQ Men and IM Eri

| | orbital signal | | superorbital signal | | negative superhump | |
|---|---|---|---|---|---|---|
| *CV* | $\omega_o$ *(c d$^{-1}$)* | *Ampl (mag)** | $\omega$ *(c d$^{-1}$)* | *Ampl (mag)** | $\omega$- *(c d$^{-1}$)* | *Ampl (mag)** |
| AQ Men | 7.0686 (3) | 0.07 | 0.263 (3) | 0.08 | 7.332 (3) | 0.04 |
| IM Eri | 6.870 (1) | 0.02 | 0.354 (7) | 0.08 | 7.226 (1) | 0.08 |

**Magnitude uncertainties are ~ 0.01 and 0.002 mag for AQ Men and IM Eri, respectively.*

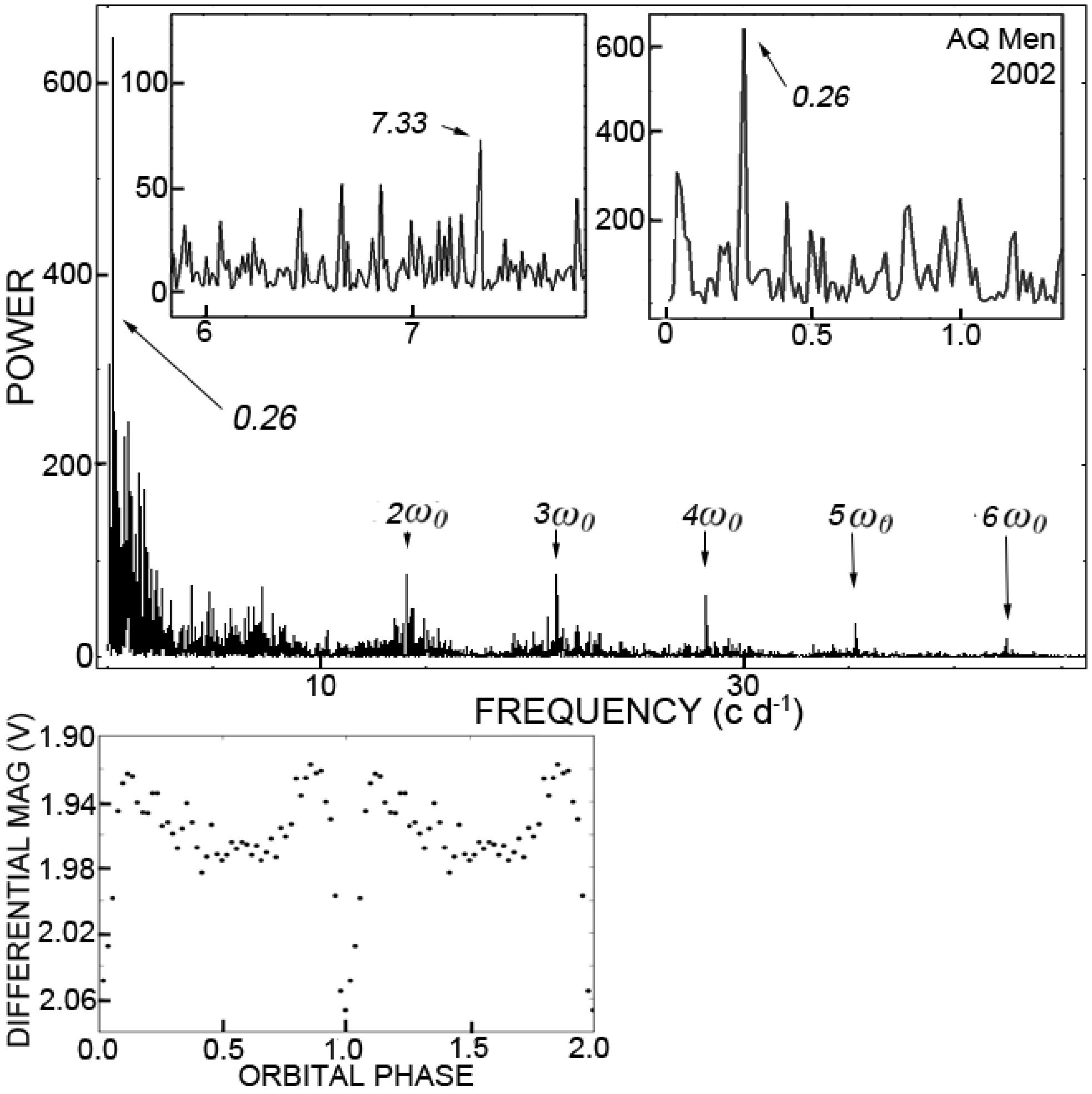

**Figure 2.** Power spectra of AQ Men in 2002. *Top*: Power spectrum of full light curve, showing the first five harmonics of a fundamental frequency of 7.0686 (3) c $d^{-1}$; *inset, left*: magnification of the power spectrum showing a weak signal at 7.332 (3) c $d^{-1}$; *inset, right*: magnification of the full power spectrum showing a strong superorbital signal at 0.263 (3) c $d^{-1}$. *Bottom*: mean orbital light curve.

*3.1.1 The orbit*

The mean orbital light curve (bottom panel of Fig. 2) looks relatively simple: a smooth modulation at a frequency of 7.0686 (3) c $d^{-1}$, with a sharp dip of ~ 0.13 mag lasting ~18 +/- 2% of the cycle. The shape and duration of this dip resemble those of most eclipsing nova-like CVs. Thus we suspect that it is an ordinary eclipse of geometric origin, likely arising from the usual cause: the transit of the secondary star across the face of the accretion disc. In this case, however, there are two significant differences. It must be a somewhat *grazing* eclipse, in order to account for the low amplitude. It is also somewhat surprising that this systematic feature in the light curve has never been clearly evident in the nightly coverage. Does it in fact recur every time around?

Yes, probably it does. From the measured precise period and the start time of the time series, we derived an eclipse ephemeris:

Minimum light = BJD 2,452,309.2781 (3) + 0.141471(8) E.

We then examined each section of light curve for the times of scheduled eclipse. Every one shows a dip, though most are heavily masked by the strong flickering. A reason for why these eclipses have been missed during past observations could be as follows: if the disc is tilted and precessing with respect to the orbital cycle, then it will present a slightly different shape to the observer each time the secondary star transits. If this reasoning is correct, then a more detailed study of the varying eclipse profile in AQ Men might provide a tool with which to probe theoretical predictions of tilted disc geometry (see §6.2).

Are our results compatible with the results of Chen et al. (2001)? Yes, they are. First, the spectra in that paper showed broad, double-peaked lines, the usual signature of high binary inclination (and especially of eclipsing CVs). Second, their radial-velocity period, though of low accuracy (0.130 (14) d), is consistent with our eclipse period. Third, their report of one isolated dip, which was not apparent on other nights, is compatible with our results (though we have many more data). The "E" symbols in Fig. 1 show the times of expected eclipse. A dip occurs at each scheduled time, though only one is sufficiently deep to be visually prominent. We thus consider it demonstrated that AQ Men displays eclipses, despite the fact that it required 33 nights of data to identify them.

### *3.1.2 The negative superhump and superorbital period*

We have established the prominence of two basic signals: an orbital frequency ($\omega_o$) and a low frequency, which we designate as N. Only two types of several-day superorbital period have been detected in CVs, and they both typically have frequencies near 0.25 c d$^{-1}$: the "short outbursts" of ER UMa-type dwarf novae (Robertson et al. 1995), and signals commonly attributed to disc precession (e.g. Patterson et al. 1997). Since no outbursts are seen and there is basically no evidence that AQ Men is a dwarf nova, we reject the former interpretation. As for the latter possibility, ten CVs are known to display signals associated with retrograde precession (see Table 2 of Montgomery (2009) and §6 for a discussion). Specifically, the signals present in these systems satisfy the following relation:

$$\omega_o + N = \omega_- \qquad 1$$

where $\omega_-$ corresponds to the frequency of a negative orbital sideband, or "negative superhump". For AQ Men, this feature would occur at at $\omega$ = 7.0686 (3) + 0.263 (3) = 7.332 (3) c d$^{-1}$.

As mentioned above, a feature in the raw power spectrum at 7.332 (3) c d$^{-1}$ (upper panel of Fig. 2, inset at left) appears marginally significant. The feature appears only slightly stronger in the prewhitened power spectrum after removing the sections of light curve with orbital phase in the range 0.90 - 1.11 (the phase range corresponding to the width of the eclipse). Yet, given the detection of the two stronger (orbital and superorbital) signals, Eq. 1 suggests that a signal at this frequency can be expected. Furthermore, the feature persists throughout the baseline of observations at roughly the same strength (this was confirmed by examining power spectra of shorter subsets of the light curve). We thus interpret it as a negative superhump.

The waveforms of the superhump and superorbital signals are shown in Fig. 3, in the left and right panels, respectively; both are approximate sinusoids. The ephemerides for these signals are:

Negative superhump:
Maximum light = BJD 2,452,309.309 (5) + 0.13646 (7) E

Superorbital signal:
Maximum light = BJD 2,452,309.26 (12) + 3.78 (5) E.

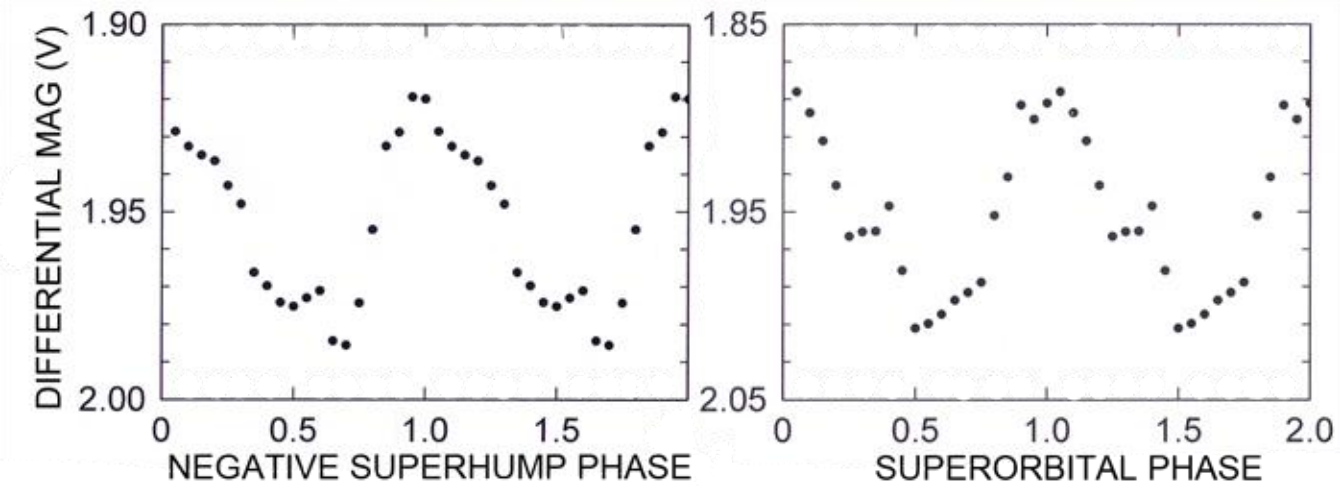

**Figure 3.** Waveforms present in the light curve of AQ Men: *Left*: negative superhump; *right*: superorbital waveform.

4. IM ERIDANI

IM Eridani (IM Eri) is a ~ 12th-magnitude object that was identified as a nova-like CV by Chen et al. (2001), who named it EC04224-2014. It has displayed a relatively constant magnitude since its discovery. Spectroscopy by Chen et al. showed shallow H absorptions with an emission reversal, and an ultraviolet excess – the defining criterion of the Edinburgh-Cape Survey. They found rapid flickering in the photometry and no convincing photometric modulations; however, radial velocity measurements revealed a possible $P_{orb}$ near 4.2 hours. On one night, the star was ~1.2 mag fainter and showed H lines predominantly in emission. These general characteristics suggest a "nova-like" interpretation, although the one-night excursion to a fainter state leaves open the dwarf-nova possibility.

*4.1 Spectroscopy*

We obtained spectra of IM Eri on 38 days between 2001 and 2011. Table 3 gives a journal of our spectroscopy, all of which is from the 1.3 m and 2.4 m telescopes at MDM Observatory. We observed the source on 37 nights during 13 different observing runs spanning 3359 days. At the 1.3-m, we used the Mark III grism spectrograph and a $1024^2$ SITe CCD, giving 2.3 Å per pixel, a resolution of 2.5 pixels, and coverage from around 4600 to 6800 Å. At the 2.4-m, we used the modular spectrograph, almost always with a $1024^2$ thinned SITe CCD, giving 2.0 Å-per-pixel, better than 2 pixel resolution, and coverage from 4300 to 7500 Å.

The top panel of Fig. 4 shows a typical average spectrum. The spectral appearance varied slightly from run to run, but IM Eri always showed a strong blue continuum and emission lines characteristic of a nova-like variable. The He I 5876 line occasionally showed weak blueshifted absorption, indicative of wind outflow. The H**α** profile was nearly Gaussian, with an emission equivalent width of typically 15 Å and a FWHM near 18 Å (800 km $s^{-1}$); H**α** occasionally showed a weak emission component extending toward the red. Synthesized V magnitudes from the spectra ranged from 11.6 to 13.0. The synthesized magnitudes are imprecise because of losses at the slit and occasional clouds, but they do show that the source was in a high state for all the spectral observations.

**Table 3 *(Sample)*.** Spectroscopy of IM Eridani

| *UT Date* | *N** | *Hour angle* | | *Tel.* |
|---|---|---|---|---|
| | | *start* | *end* | |
| 2001 Nov 20 | 2 | +01:24 | +01:35 | 2.4m |
| 2001 Nov 21 | 2 | +02:03 | +02:21 | 2.4m |
| 2001 Nov 22 | 10 | -02:08 | +01:52 | 2.4m |
| 2001 Nov 24 | 8 | -02:40 | +03:26 | 2.4m |
| 2001 Dec 19 | 2 | -01:42 | -01:34 | 1.3m |

** N = Number of frames*

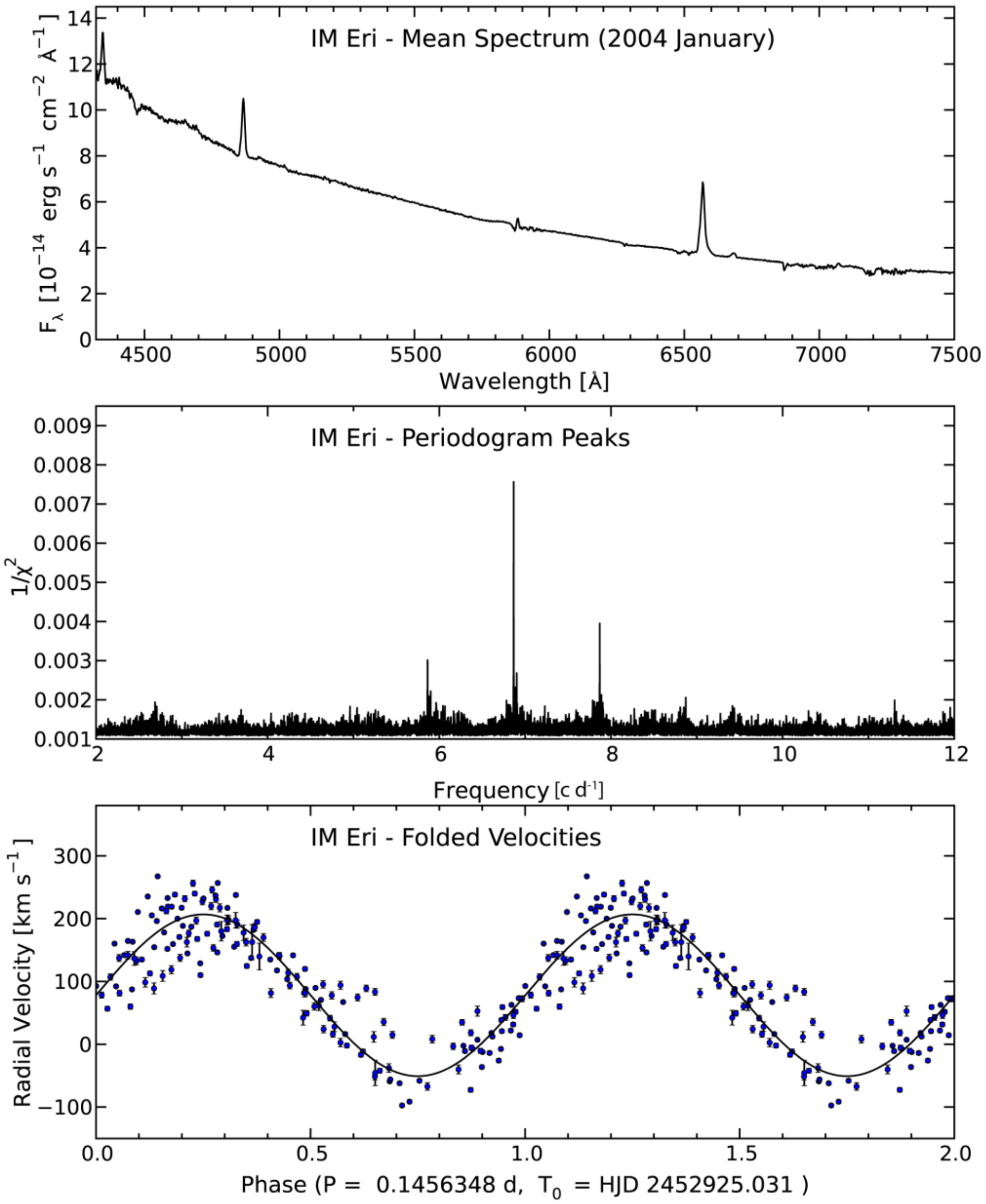

**Figure 4**. *Top:* Mean spectrum of IM Eri taken 2004 January. *Middle:* Result of a period search of the H**α** radial velocities. The prominent spike just below 7 c $d^{-1}$ corresponds to the adopted period; the candidate periods separated by +/- 1 c $d^{-1}$ are much less likely. *Bottom*: H**α** radial velocites folded on the indicated ephemeris. The best-fitting sinusoid has a semi-amplitude K = 129 (9) km $s^{-1}$ and an average velocity of 78 (6) km $s^{-1}$.

Table 4, which is available in the online version of this article, gives the complete list of H**α** radial velocities. We measured these using a convolution method (Schneider & Young 1980, Shafter & Szkody 1984). The convolution function consisted of two Gaussians, one positive and the other negative in velocity with respect to line center, each with FWHM 8 Å and separated by 26 Å. This emphasized the steep sides of the line profile. Analyzing these with a "residual-gram" (Thorstensen et al. 1996) yielded a period of 0.1456348 (4) d, with no significant ambiguity in cycle count on any time scale. The middle panel of Fig. 4 shows this analysis, and the bottom panel shows all the data folded on this period, with the best-fitting sinusoid superposed.

**Table 4 *(Sample)*.** IM Eridani: H**α** Radial Velocities

| *BJD (2400000+)* | *velocity (km/s)* | *sigma (km/s)* |
|---|---|---|
| 52233.8836 | 129 | 2 |
| 52233.8911 | 172 | 2 |
| 52234.9102 | 180 | 15 |
| 52234.9231 | 140 | 21 |
| 52235.7332 | -7 | 2 |
| 52235.7366 | 22 | 3 |

*4.2 Photometry*

We obtained 90.5 hours of photometry on IM Eri during 21 nights over 93 days in 2002, and 144.5 hours during 26 nights over 32 days in 2012. Table 1 contains the observing log. Most of the data were obtained in V light. The mean magnitude of the object was V = 11.4 (2) in 2002 and V = 11.7 (2) in 2012. The visual appearance of the light curves was similar during the two years, and during each night. Because the 2012 coverage was more extensive and better distributed in time, most of our analysis focused on those observations. The top and bottom panels of Fig. 5 show the light curves obtained in 2002 and 2012, respectively.

*4.2.1 2012*

The bottom frame of Fig. 5 shows the full 32-day light curve of 2012, and inset is a representative ~ 7-hour window. Most of the data were from the Americas, but several nights from South Africa were sufficient to discriminate between the signals and the daily aliases. We used the full 32-day light curve to determine period values. The top panel of Fig. 6 shows the power spectrum in the frequency range of 5 – 10 c $d^{-1}$. The power spectrum is dominated by a signal at 7.226 (1) c $d^{-1}$, which persists at roughly constant amplitude (0.08 (1) mag) over the baseline of observations. The bottom panel of Fig. 6 shows the power spectrum of the full light curve in the frequency range: 0 – 2.2 c $d^{-1}$, in which a feature is present at 0.354 (7) c $d^{-1}$ with an amplitude of 0.08 (1) mag.

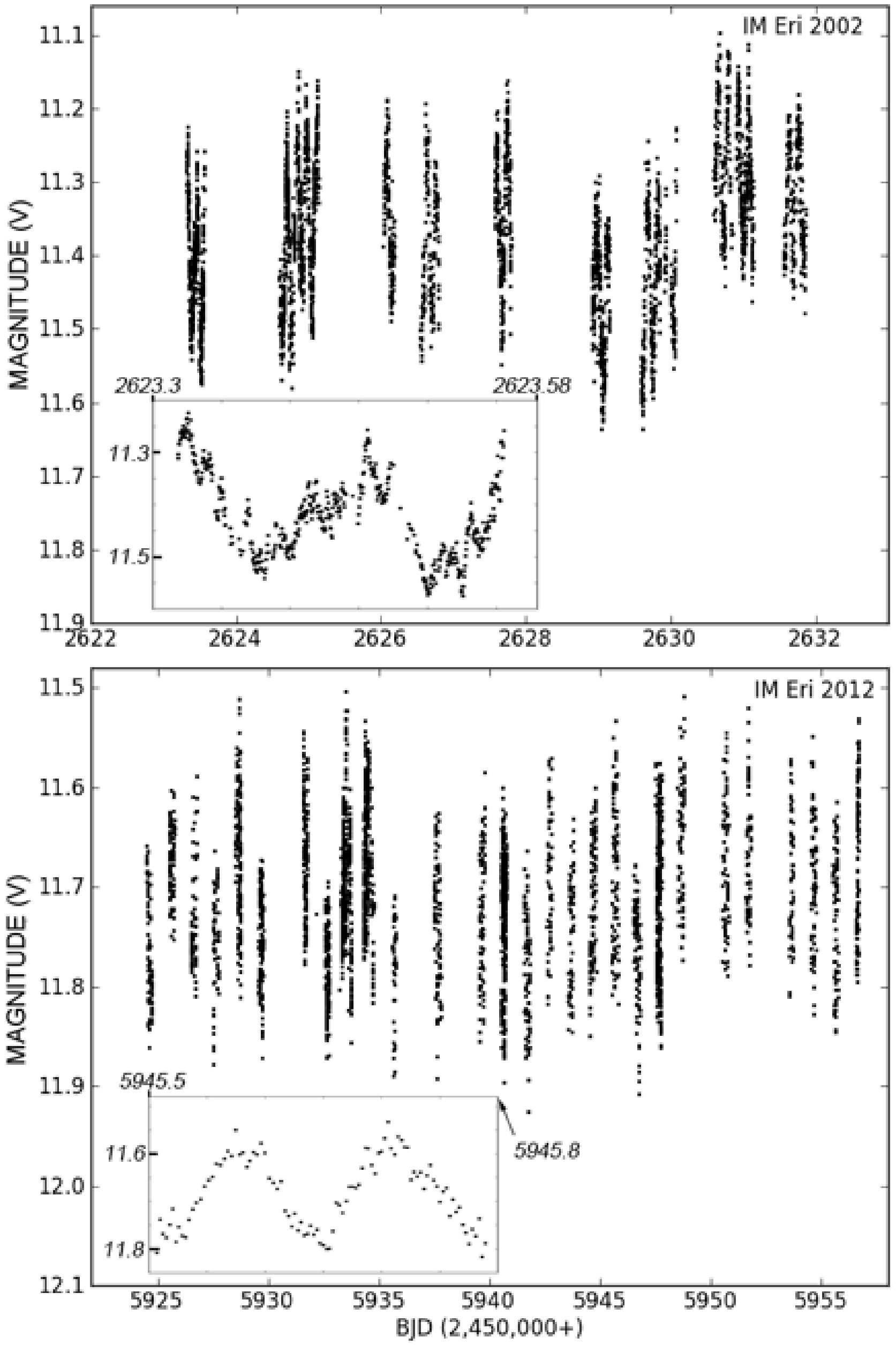

**Figure 5.** *Top*: Light curve of IM Eri in 2002; *inset:* magnification of a 6.7-hour observations window. *Bottom:* Light curve of IM Eri in 2012; *inset:* magnification of a 7.2-hour observations window. The V magnitudes have a systematic calibration error of ~ 0.3 mag.

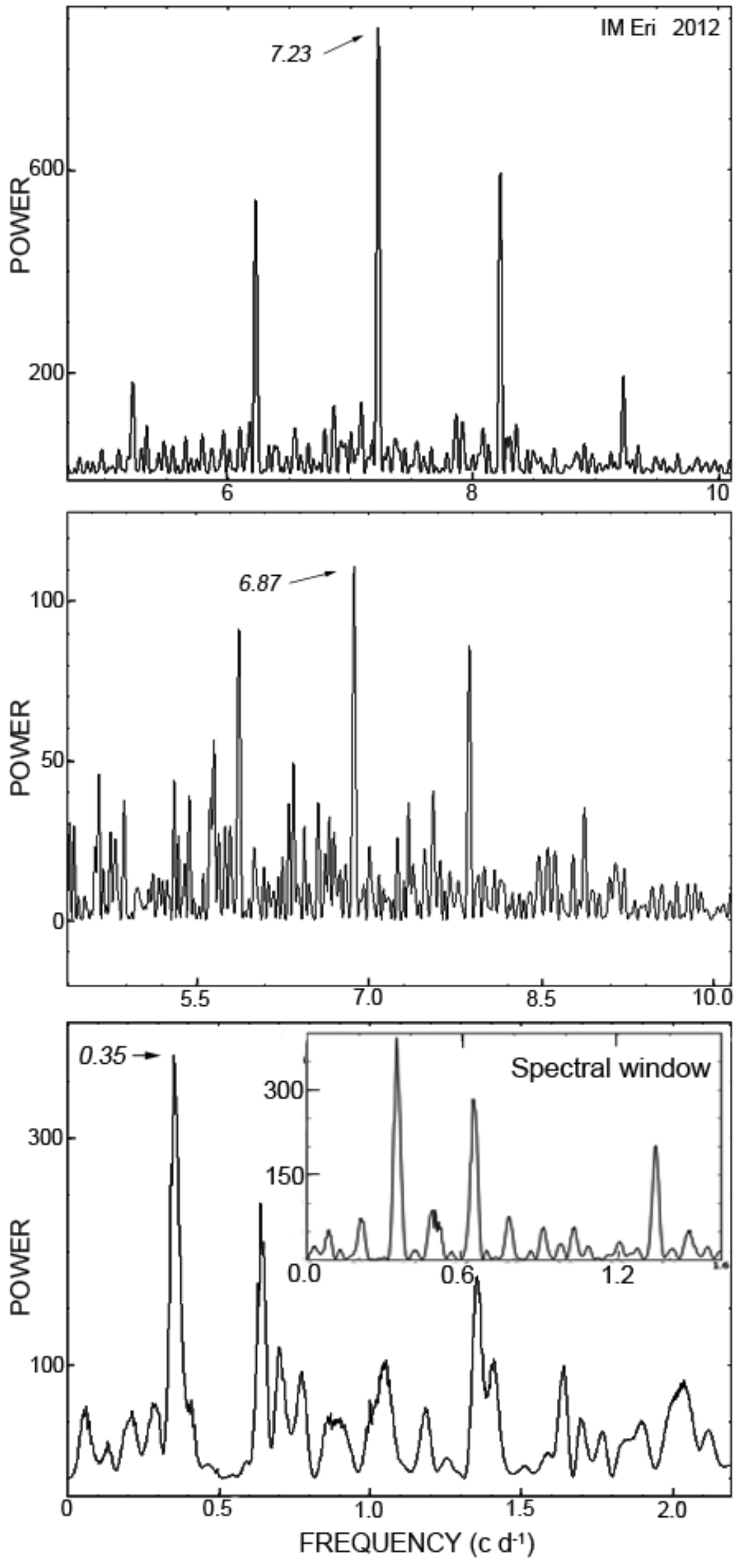

**Figure 6.** Power spectrum of IM Eri in 2012. *Top*: the frequency range showing the prominent superhump signal at 7.226 (1) c $d^{-1}$. *Middle*: the same frequency range after the subtraction of the superhump, showing the orbital signal at 6.870 (1) c $d^{-1}$. *Bottom*: the frequency range of the original power spectrum showing the superorbital signal at 0.354 (7) c $d^{-1}$.

Both signals are flanked by aliases. The faster signal has merely the familiar +/- 1 c $d^{-1}$ structure. But with observations of ~1-day spacing, a low-frequency signal (at $\omega \cong 0.35$ c $d^{-1}$) will be aliased with signals at (1- ω) and (1+ ω) c $d^{-1}$, and possibly also at 1 c $d^{-1}$ if differential extinction is significant. This is important to recognize here, because what appears as "noise" in the low-frequency power spectrum is actually dominated by aliases of the low-frequency signal. The spectral window of the signal at 0.35 c $d^{-1}$ is inset for comparison in the bottom panel of Fig. 6.

To search for an orbital signal, we subtracted the superhump from the full light curve and obtained the prewhitened power spectrum, which is shown in the middle panel. A signal is present at 6.870 (1) c $d^{-1}$, which is consistent with the radial-velocity period. We interpret it as the orbital signal. The best-fit sinusoid has an amplitude of 0.020 (4) mag.

Thus we have a familiar triad of frequencies that satisfies Eq. 1: the orbital frequency at 6.86649 (2) c $d^{-1}$, (the precise radial-velocity period), a negative superhump at 7.226 (1) c $d^{-1}$, and a superorbital signal at 0.354 (7) c $d^{-1}$. In Fig. 7 we show the mean waveforms of these signals. The extrema satisfy the ephemerides:

Orbit:
Minimum light = BJD 2,455,924.522 (2) + 0.1456346 (2) E
Negative superhump:
Maximum light = BJD 2,455,924.532 (2) + 0.13841 (3) E
Superorbital signal:
Maximum light = BJD 2,455,925.85 (7) + 2.809 (15) E.

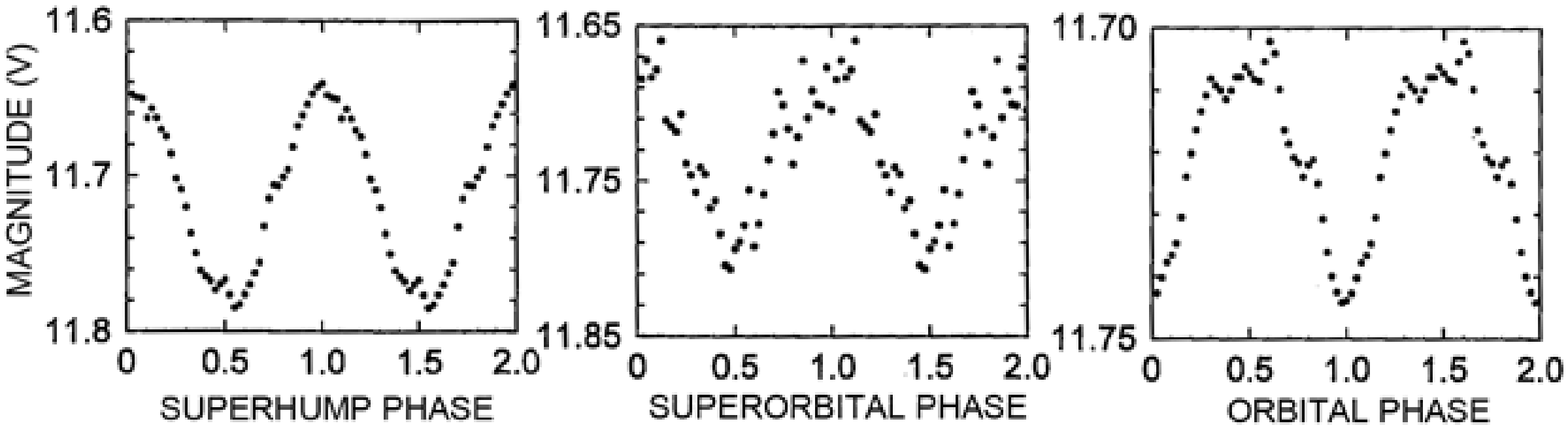

**Figure 7.** Waveforms present in the light curve of IM Eri in 2012. *Left*: superhump; *middle*: superorbital waveform; *right*: orbital waveform.

*4.2.2 2002*

The baseline of observations in 2002 spanned 93 days, but the time coverage was not optimal (truncated JD = 538-9, 602-6, 623-31). The top panel of Fig. 5 shows the full light curve and a representative ~ 7-hour window (inset). Only the last nine days were sufficiently calibrated (between telescopes) to search for the superorbital signal. Consequently, the results of this season were more poorly constrained than those of 2012. The power spectrum created from a light curve of the final nine days yielded significant detections at 0.34 (3) and 7.22 (1) c $d^{-1}$, with amplitudes of 0.05 (2) and 0.09 (2) mag, respectively. To search for an orbital period at the spectroscopic value of 6.87 c $d^{-1}$, we removed the strong 7.22 c $d^{-1}$ feature from the light curve. The prewhitened power spectrum showed a possible feature at 6.862 (10) c $d^{-1}$, but it was too weak to count as a detection of the orbital signal. In summary, all the detections made from these data are consistent with the results of the more extensive 2012 analysis.

## 5. TESTS OF THE SUPERORBITAL FEATURES

Sometimes, two periodicities that are present in a light curve can interact so as to produce an artifact in the power spectrum. The artifact can appear significant even though it does not represent a physically meaningful feature of the system under study. This phenomenon is not usually a problem for orbital and orbital sideband signals, because their periods are short compared to the durations of our photometry campaigns. Long-period signals, however, can be hard to distinguish from a CV's innate erratic variability. Hence, we performed two tests upon the superorbital features in AQ Men and IM Eri, to establish their independence from the orbital and superhump signals.

First we created a light curve comprised of two sinusoids, one representing the orbit and the other the superhump, to determine whether the interaction of these signals alone could reproduce power at the superorbital frequency. We sampled the sinusoids using the sampling rates and observing windows of the real data and added Gaussian noise to simulate the signal-to-noise level in the power spectrum of the real data. As a second test, we subtracted the orbital and superhump signals from the light curve and examined the resulting power spectrum for any feature remaining at the superorbital frequency.

The results of these two tests for the 2012 IM Eri data are shown in Figures 8 and 9, respectively. Fig. 8 shows the power spectra of the real data (top) and the two model sinusoids (bottom): power at the superorbital frequency of 0.354 c d$^{-1}$ is not reproduced by the model. Fig. 9 shows the result of the second test, comparing the power spectrum of the full light curve (top) to the prewhitened power spectrum (bottom). The superorbital feature remains after the removal of the faster signals. Thus, both tests indicate that the superorbital feature is an independent signal in this system. Results of these tests also confirmed the independence of the superorbital signal in AQ Men.

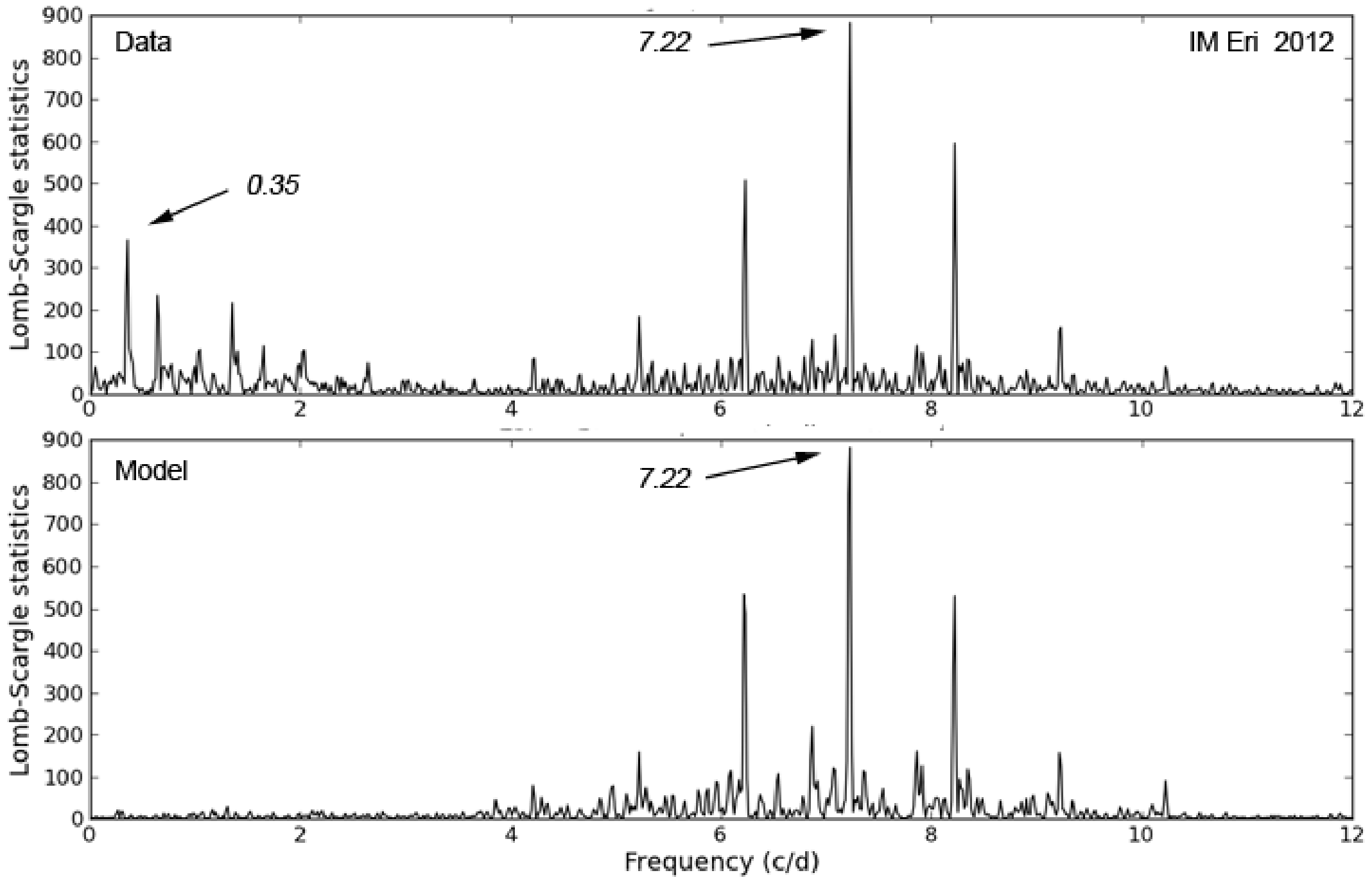

**Figure 8.** *Top:* power spectrum of IM Eri in 2012, showing the superhump at 7.226 (1) c d$^{-1}$ and superorbital signal at 0.354 (7) c d$^{-1}$. *Bottom:* power spectrum created from two model sinusoids representing the orbital and superhump signals. No power is present at the superorbital frequency.

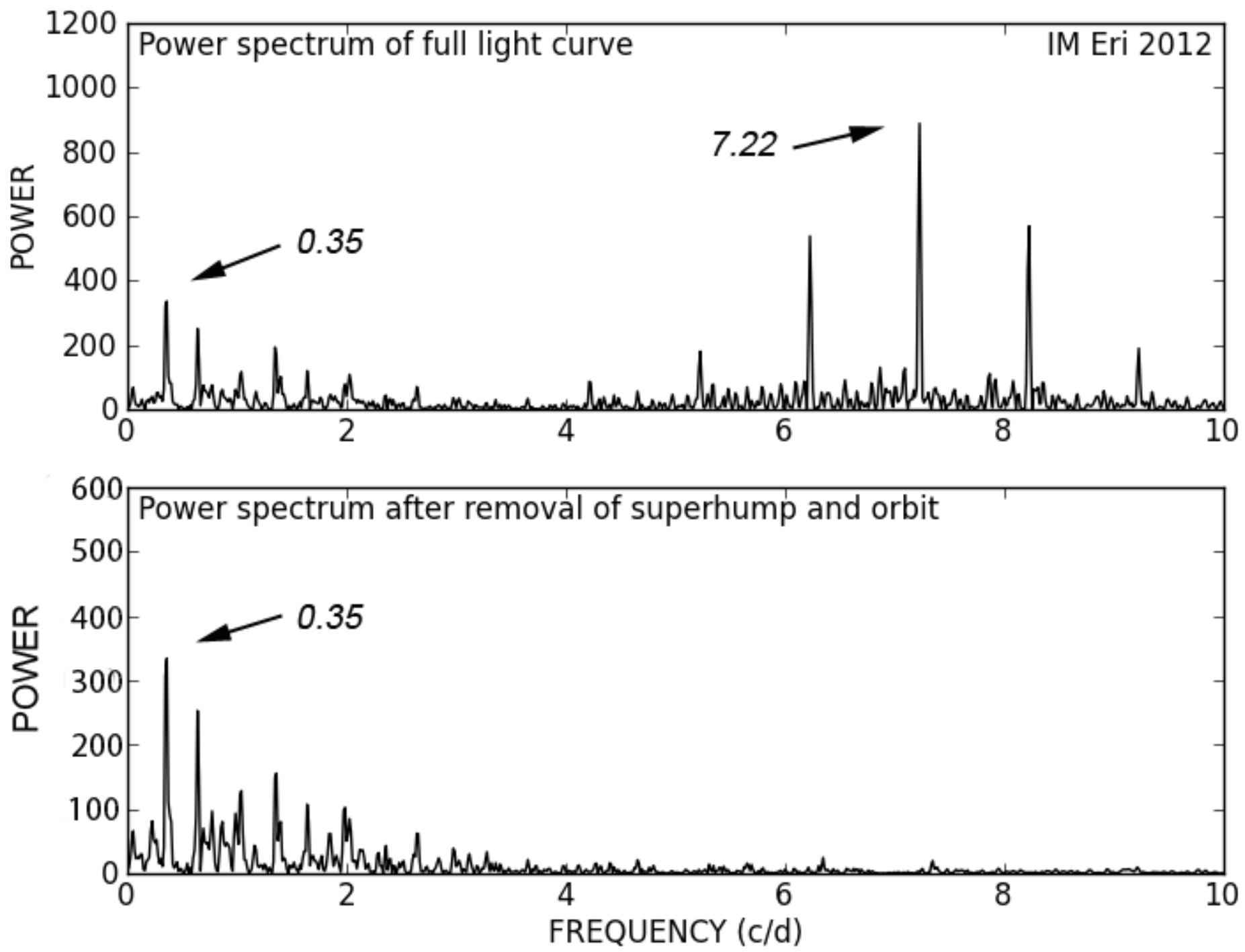

**Figure 9.** *Top panel*: power spectrum of IM Eri in 2012, showing the superhump and superorbital signals. *Bottom*: prewhitened power spectrum after the removal of the superhump and orbital signals. The statistical significance of the feature at the superorbital frequency slightly increases after the subtraction.

## 6. SUPERORBITAL SIGNALS IN CVS and LMXBs

### *6.1 Summary of the theory*

The most commonly accepted model for simultaneous superorbital and negative superhump periods in CVs and LMXBs involves a tilted accretion disc. The line of nodes of the disc retrogradely precesses, and the interaction of the precession and orbital motions produces the negative superhump period. The origin of disc tilt in LMXBs is likely to be irradiation by the primary compact object (Pringle 1996, Wijers & Pringle 1999, Maloney & Begelman 1997): Pringle (1996) showed that a sufficiently high central radiation pressure force will render an initially flat disc unstable to tilting, and that the primary stars in LMXBs meet this criterion.

In CVs, irradiation by the primary is probably insufficient to produce tilt (Petterson 1977, Iping & Petterson 1990), and there exists no consensus on an alternative cause. Possibilities include the magnetic field of the secondary (Murray et al. 2002) and lift (theory: Montgomery & Martin 2010; simulation: Montgomery 2012). Regardless of the cause of tilt, *if* a minor tilt perturbation on a disc in a CV can be incited, two reinforcing effects may follow. First, the two hemispheres of the secondary may experience variable asymmetric irradiation by the primary, due to the disc's shadow – which may vary the flux of the mass transfer stream (Smak 2009a, 2009b; Barrett et al. 1988). Secondly, the stream will have a nonzero vertical velocity component with respect to the tilted disc, and thus will not strike the disc edge-on. Rather, infalling matter will encounter the disc face, imparting greater force upon one side over the other, thus possibly reinforcing tilt. The negative superhump modulation results from energy dissipation at varying depth of the stream-disc impact point in the potential well of the primary (Smak 2009b).

Hydrodynamic simulations can reproduce negative superhump observations by attributing the observed modulation to the transit of the bright spot across the face of a tilted disc (Wood & Burke 2007; these authors also share elucidating animations: http://astro.fit.edu/wood/visualizations.html). The modulation at the retrograde precession period might be due to varying projected surface area of the disc (Patterson et al. 2002).

### *6.2 Our observations in context*

Our observations of AQ Men and IM Eri bring the number of CVs known to display simultaneous negative superhumps and superorbital periods to 12 – a significant increase in sample size. In Table 5 we tabulate these 12 systems, eight of which are reliable detections and four of which are likely. The "likely" four objects, listed with asterisks, have plausible evidence for such a signal, usually in the form of a folded light curve, but with no published power spectrum – which is needed to characterize the noise level at such low frequencies. All 12 objects show negative superhump and orbital periods, which, together with the value of the superorbital period, satisfy Eq. 1. In addition to this small group, there are another ~ 15 CVs that show negative superhump periods, which have been tabulated in Table 2 of Montgomery (2009) – along with *calculated* retrograde precession periods. For almost all of these objects, however, no periodic-signal search at low frequency has been performed. That is: neither the presence of a precession period nor a lower limit on non-detections have been made.

**Table 5.** Superorbital Periods Observed in CVs, Derived from Photometry and Accompanied by Negative Superhumps

| *CV* | $P_{orb}$ *(d)* | $P_{-}$ *(d)* | $P_{superorb}$ *(d)* |
|---|---|---|---|
| BK Lyn[1] | 0.07494 (1) | 0.07279 (1) | 2.544 (12) |
| RX 1643+34[2] | 0.12056 (2) | 0.11696 (7) | 4.05 (13) |
| V442 Oph | 0.12435 (7)[3] | 0.12090(8)[2] | 4.37 (15)[2] |
| AH Men[4] | 0.12721 (7) | 0.1230 (1) | 3.68 (3) |
| TT Ari | 0.1375511(1)[5] | 0.1328 (2)[6] | 3.8 (2)[7]* |
| AQ Men[8] | 0.141471 (6) | 0.13639 (5) | 3.80 (4) |
| V1193 Ori[9] | 0.1430 (1) | 0.1362 (1) | 2.98 (8)* |
| V751Cyg[10] | 0.1445 (2) | 0.1394(1) | 3.94(6) |
| IM Eri[8] | 0.1456348 (4) | 0.13839 (2) | 2.82 (6) |
| PX And[11] | 0.1463527 (1) | 0.1415 (3) | 4.8 (4)* |
| SDSS J0407-06[9--12] | 0.17017 (1) | 0.166 (1) | 5.3 (7)* |
| TV Col | 0.228599 (1)[13] | 0.21667 (10)[14] | 4.0 (8)[15] |

* Period likely but not established beyond doubt.

REFERENCES. (1) Patterson et al. 2012; (2) Patterson et al. 2002; (3) Hoard et al. 2000; (4) Patterson 1995; (5) Wu et al. 2002; (6) Kraicheva et al. 1999; (7) Udalski 1988; (8) This paper; (9) Ak et al. 2005a; (10) Patterson et al. 2001; (11) Stanishev et al. 2002; (12) Ak et al. 2005b; (13) Augusteijn et al. 1994; (14) Retter et al. 2003; (15) Motch 1981.

Secondly, we have discovered AQ Men to be a member of this small group of objects – which also shows grazing eclipses. If our interpretation of the variable orbital waveform in AQ Men is correct, that is: if it is a manifestation of the disc's tilt (§3.1.1), then detailed study of these shallow eclipses might offer an opportunity to explore the disc geometry. In particular: is the variable shape of the eclipse profile itself periodic? Such a study may be accomplished via more extensive photometry campaigns that can adequately define the shapes of individual eclipses. We shall make specific remarks in §6.3.

*6.3 Questions*

We shall mention three open questions that may be illuminated by further detailed photometric studies of superorbital periods in CVs, and by searches of CVs that are likely to display these periods.

*6.3.1 Are retrograde precession periods more common in LMXBs than in CVs?*

The number of LMXBs that have displayed superorbital periods associated with retrograde disc precession may be as many as 25 (Kotze & Charles 2012) out of about 100 known (Ritter & Kolb 2012). (In addition to Her X-1, other well studied systems are SS 433 (Cherepashchuk 2002), LMC X-4 (Lang et al. 1981; Paul & Kitamoto 2002), and SMC X-1 (Wojdowski et al. 1998)). In contrast, of the ~ 100 CVs that we estimate have been searched for superorbital periods via time series photometry, only the twelve listed in Table 5 have had detections. Why the disparity?

Feasibility is likely to be a significant reason. To find signals with periods of days, one usually requires dense observations over weeks. While some X-ray telescopes have surveyed the sky continually for years (e.g. the High Energy Astronomy Observatory – 1 (HEAO-1), Rossi X-ray Timing Explorer (RXTE)), obtaining sufficient time on ground-based telescopes usually requires the dedicated work of coordinated groups – and the cooperation of the weather. It is also possible that disc tilt is intrinsically more common in LMXBs, given that a likely tilt mechanism in CVs has yet to be firmly identified. But this possibility is mere speculation on our part. It is a fascinating issue for future study.

*6.3.2 Long-term stability of a tilted disc?*

Negative superhumps and superorbital periods associated with retrograde precession are not strictly persistent features of the systems in which they have been detected. Generally, these signals persist throughout an observing season of several months' duration, are not detected a year or so later during an observing season of comparable length, and reappear several years after that (Patterson et al. 1997). If understood within the context of tilted-disc geometry, then these observations indicate: 1) a disc can remain stably tilted over a timescale of months but probably not years; 2) after the disc has re-aligned with the orbital plane and maintained this alignment for a year or so, it returns to its previous tilted state. Why might this occur? It is possible that phenomena associated with *positive* superhumps play a role (e.g. Wood et al. 2011). More extensive monitoring of these CVs may provide tighter constraints on the time frames in question.

*6.3.3 Varying tilt of the disc?*

The eclipses of AQ Men vary greatly in depth. Even *consecutive* eclipses vary greatly, and this suggests that erratic flickering contributes significantly to the variability. But the shallowness of the average eclipse also requires that the eclipses be grazing, and this produces a large asymmetry: the back of the disk – where "back" is in reference to the observer – is never eclipsed, while the front is. With a more extensive observational campaign, we might study whether the variable shape of the eclipse profile is itself periodic. Such a study appears to be eminently feasible for a star with declination ~ -80˚, and we hope to use it to constrain the disc geometry.

One possible application of studying the varying eclipse profile in AQ Men is a test of simulations done by Montgometry (2012). Specifically: does the orientation of a tilted disc with respect to the mass transfer stream influence the depth of the stream-disc impact point in the potential well of the primary? Montgomery (2012) showed that, for a disc tilted at 5˚ with respect to the orbital plane, the stream flows over the disc rim for ~ half an orbit and then under for ~half an orbit, and that as the disc precesses, the location on the disc rim where the stream transitions from over-the-disc to under-the-disc moves in the retrograde direction. This possibility might be investigated via a study of the superhump strength as a function of the phase of the superorbital signal – since the superhump modulation is thought to be due to varying depth of the stream-disc impact point in the potential well of the primary.

We note that a few detailed optical studies of variability linked to superorbital timescales have been done with LMXBs, and have illuminated questions regarding the systems' geometry. Deeter et al. (1976), for example, found optical modulations in Her X-1 repeating on the 35-day retrograde precession period, and attributed them to the uniformly-changing orientation of luminous parts of the system.

## 7. SUMMARY

AQ Men and IM Eri have displayed orbital, superorbital, and negative superhump signals, which can be understood within the context of an accretion disc that is tilted out of the orbital plane and whose line of nodes is precessing in a retrograde direction with respect to the orbit. This identification brings the number of CVs with firm detections of signals associated with retrograde precession to twelve. Furthermore, AQ Men has been found to display grazing eclipses, which may provide a unique opportunity to study the variable disc geometry in detail. Finally, more long-term time series photometry of CVs that are likely to display superorbital signals – such as systems with negative superhump detections – may constrain models of tilted accretion discs for this class of object.

## ACKNOWLEDGEMENTS

For their support of this research, we thank the Mount Cuba Astronomical Foundation, and the National Science Foundation through grants AST-0908363 and AST-1211129. We also thank the scientific editor at MNRAS for comments that dramatically improved the quality of this paper.